\documentclass[aps,preprint,amsmath]{revtex4}

\usepackage{graphics,graphicx,amssymb,amsmath}
\usepackage{setspace}

\begin{document}

\title{Fractionalized excitations in the spin liquid state of a kagom\'{e} lattice antiferromagnet}

\author{Tian-Heng~Han$^{1,\dag}$, Joel S.~Helton$^{2}$, Shaoyan~Chu$^{3}$, Daniel G.~Nocera$^{4}$, Jose A.~Rodriguez-Rivera$^{2,5}$, Collin~Broholm$^{2,6}$ \& Young S.~Lee$^{1}$}

\affiliation{$^{1}$Department of Physics, Massachusetts Institute of Technology, Cambridge, Massachusetts 02139, USA}
\affiliation{$^{2}$NIST Center for Neutron Research, National Institute of Standards and Technology, Gaithersburg, Maryland 20899, USA}
\affiliation{$^{3}$Center for Materials Science and Engineering, Massachusetts Institute of Technology, Cambridge, Massachusetts 02139, USA}
\affiliation{$^{4}$Department of Chemistry, Massachusetts Institute of Technology, Cambridge, Massachusetts 02139, USA}
\affiliation{$^{5}$Department of Materials Science and Engineering, University of Maryland, College Park, Maryland 20742, USA}
\affiliation{$^{6}$Institute for Quantum Matter and Department of Physics and Astronomy, The Johns Hopkins University, Baltimore, Maryland 21218, USA.}

\date{\today}

\maketitle

\small

New physics can emerge in magnetic materials where quantum fluctuations are enhanced due to reduced dimensionality and strong frustration. One long sought example is the resonating-valence-bond (RVB) state,\cite{Anderson} where atomic magnetic moments are strongly correlated but do not order or freeze even in the limit of $T \rightarrow 0$.\cite{Balents} The RVB ground state does not break conventional symmetries, such as lattice translation or spin-rotation.  The realization of such a quantum spin liquid in two-dimensions would represent a new state of matter. It is believed that spin liquid physics plays a role in the phenomenon of high-$T_C$ superconductivity,\cite{PLee_review} and the topological properties of the spin liquid state may have applications in the field of quantum information\cite{Ioffe}.  We present neutron scattering measurements of the spin excitations on single crystal samples of the spin-1/2 kagom\'{e} lattice antiferromagnet ZnCu$_3$(OD)$_6$Cl$_2$ (also called herbertsmithite).  Our observation of a spinon continuum in a two-dimensional magnet is remarkable first.  The results serve as a key fingerprint of the quantum spin liquid state in herbertsmithite.

\

A hallmark feature of quantum spin liquids is the presence of {\em deconfined spinons} as the fundamental excitation from the ground state.\cite{Sachdev1992} Spinons are $S=\frac{1}{2}$ quantum excitations, into which conventional spin-wave excitation with quantum number $S=1$ fractionalize. In one dimension, this phenomenon is well established for the spin-$\frac{1}{2}$ Heisenberg antiferromagnetic chain, where spinons may be thought of as magnetic domain boundaries that disrupt N\'{e}el order and are free to propagate away from each other \cite{Tennant}. In two-dimensions, the character of spinon excitations is less clear.  First, the existence of two dimensional magnets with a quantum spin liquid ground state is still a matter of great debate.  Second, the various spin liquid states which are possible in theory give rise to a variety of spinon excitations, which can form a spinon Fermi surface or have a Dirac quasiparticle spectrum.\cite{Ran}

\

The spin-$\frac{1}{2}$ kagom\'{e} lattice Heisenberg antiferromagnet has long been recognized as a promising system in which to search for quantum spin liquid states, as the kagom\'{e} network of corner-sharing triangles frustrates long-range magnetic order\cite{Elser, Marston,Sachdev1992,Lecheminant}.  We have devised synthetic methods to deliver the compound herbertsmithite (ZnCu$_{3}$(OH)$_{6}$Cl$_{2}$) where the spin-$\frac{1}{2}$ Cu$^{2+}$ moments are arranged on a structurally perfect kagom\'{e} lattice\cite{Shores}, with nonmagnetic Zn$^{2+}$ ions separating the planes.  Whereas this material typically contains a small percentage of excess Cu$^{2+}$ ions ($\sim 5\%$ of the total) which substitute for Zn$^{2+}$ ions in the interlayer sites, the kagom\'{e} planes only contain Cu$^{2+}$ ions.\cite{Freedman}   Measurements on powder samples\cite{Helton,Mendels,deVries} indicate strong antiferromagnetic super-exchange ($J \, \approx$~17~meV~or~200~K) and the absence of long-range magnetic order or spin freezing down to temperatures of $T=0.05$~K.  A small anisotropy is observed in the bulk magnetic properties, indicating a Dzyaloshinskii-Moriya interaction as well as an easy-axis exchange anisotropy,\cite{Zorko,Han2} both of order $J/10$.  Despite these small imperfections, the nearest neighbor Heisenberg model on a kagom\'{e} lattice is still an excellent approximation of the spin Hamiltonian for herbertsmithite.  This is especially important, since recent calculations on record lattice sizes indicate that the ground state of this model is, in fact, a quantum spin liquid.\cite{White}  Thus, experiments to probe the spin correlations in herbertsmithite are all the more urgent.

\

Towards this end, we have recently succeeded in developing a new technique for the growth of large, high quality, single crystals of herbertsmithite,\cite{Han} and small pieces have been used in studies involving local-probes,\cite{Imai,Ofer} anomalous x-ray diffraction,\cite{Freedman} susceptibility,\cite{Han2} and Raman scattering\cite{Wulferding}. In this letter we report inelastic neutron scattering measurements on a large deuterated single crystal sample of herbertsmithite. Inelastic neutron scattering experiments were performed using the Multi-Axis Crystal Spectrometer (MACS) at the NIST Center for Neutron Research. A pumped helium cryostat was used to cool the sample to $T=1.6$~K.  The final analyzed neutron energy was either $E_{f}=5.1$~meV or $E_{f}=3.0$~meV, for energy resolutions of 0.21~meV (half-width at half-maximum) and 0.08~meV, respectively.  The neutron scattering cross section is directly proportional to the dynamic structure factor $S_{tot}(\vec{Q},\,\omega)$, which includes both the nuclear and magnetic signals. The magnetic part, $S_{mag}(\vec{Q},\,\omega)$, can be obtained by subtracting the nuclear scattering as described in the supplementary information. After calibration with respect to a vanadium standard, the structure factors are plotted in absolute units.

\

Contour plots of $S_{tot}(\vec{Q},\,\omega)$ are shown in Figures~1(a)-(c) for $T=1.6$~K and three different energy transfers $\hbar\omega$. Figure~1(a) shows $\hbar\omega=6$~meV data. Surprisingly, the scattered intensity is exceedingly diffuse, spanning a large fraction of the hexagonal Brillouin zone.  A similar pattern of diffuse scattering is observed for $\hbar\omega=2$~meV (Fig.~1(b)). The diffuse nature of the scattering at a temperature that is two orders of magnitude below the exchange energy scale $J$ is in strong contrast to observations in non-frustrated quantum magnets.  The $S=1/2$ square lattice antiferromagnet La$_2$CuO$_4$ develops substantial antiferromagnetic correlations for $T < J/2$ \cite{Birgeneau}, where the low energy scattering is strongly peaked near the $(\pi,\pi)$ point in reciprocal space.  In herbertsmithite, the scattered intensity is not strongly peaked at any specific point, and this remains true for all energies measured from $\hbar\omega = 0.25$~meV to 11 meV.

\

The observed $\vec{Q}-$dependence of the scattered intensity provides important information on the ground-state spin correlations.  The scattering in reciprocal space has the shape of broadened hexagonal rings (or donuts) centered at $(0,0,0)$ and $(2,0,0)$-type positions. All of the scans that we have performed from $\hbar\omega=1.5$~meV to 11~meV show similar patterns for the scattered magnetic intensity.  The energy-integrated dynamic structure factor over the integration range $1 \leq \hbar\omega \leq 9$~meV is plotted in Fig.~1(d).  This quantity serves as an approximation of the equal-time structure factor. For comparison, a calculation of the equal-time structure factor for a collection of uncorrelated nearest neighbor singlets on a kagom\'{e} lattice is shown in Fig.~1(e).  To first approximation, the observed magnetic signal resembles this calculation.  Therefore, the ground state wave function of herbertsmithite has a large component resembling randomly arranged nearest neighbor singlets, consistent with a short-range RVB state\cite{Anderson,Kivelson,White}.  However, it is also clear that the data has sharper features than the model calculation.  Thus, the spin-spin correlations in herbertsmithite extend beyond nearest neighbors, as further discussed below.  The intensity in Fig.~1(e) corresponds to 1/8 of the total moment sum rule\cite{Lovesey}.  For the data, the integrated intensity up to $\hbar\omega = 11$~meV corresponds to 20(3)\% of the total moment.

\

At the lowest measured energy transfers, we observe additional features in the pattern of magnetic scattering.  Figure~1(c) depicts the intensity contour plot for $\hbar\omega=0.75$~meV showing additional broad peaks centered at $(1,0,0)$ and equivalent positions.  The $(1,0,0)$ position does not correspond to a nuclear Bragg position for this crystal structure.  Additional scans taken with $\hbar\omega$ between 0.25~meV and 1~meV confirm that this feature is generic to the low energy transfers.  This peak is likely influenced by the weakly coupled Cu$^{2+}$ ions on the interlayer Zn$^{2+}$ sites, which are believed to affect the low energy scattering.\cite{Helton2}

\

The overall insensitivity of the pattern of the scattering to energy transfer is another remarkable feature of the data.  Conventional spin-wave excitations take the form of sharp surfaces of dispersion in $(\vec{Q},\omega)$-space \cite{Matan}.  Here, no surfaces of dispersion are observable in the low-temperature data.  The $\hbar\omega$ versus $\vec{Q}$ dependence of $S_{mag}(\vec{Q},\,\omega)$ is plotted in Figure 2 for two high-symmetry directions in reciprocal space: the $(H~0~0)$ direction in Fig.~2(a) and the $(H~H~0)$ direction in Fig.~2(b).  These directions are indicated by thick black lines in Fig.~2(d). These plots show that the spin excitations form a broad, continuous band (or a continuum), extending up to the highest measured energy of 11 meV.  This is direct evidence that the excitations are fractionalized, forming a two-spinon continuum in this two-dimensional antiferromagnet.

\

In Fig.~2(c) and its inset, the energy dependences of $S_{tot}(\vec{Q},\,\omega)$ and $S_{mag}(\vec{Q},\,\omega)$ are plotted for high symmetry $\vec{Q}$-positions as indicated in the reciprocal space map in Fig.~2(d). The scattered signal is rather flat for $2 \leq\hbar\omega\leq 10$~meV but increase significantly with decreasing energy-transfer below $\hbar\omega = 1.5$~meV.  Clearly, there is no indication of a spin-gap down to $\hbar \omega = 0.25$~meV at the measured reciprocal-space positions.

\

The magnetic intensity can be plotted as one-dimensional ``line-scans'' along specific direction in reciprocal space. In Fig.~3(a), $S_{mag}(\vec{Q},\,\omega)$ is shown along the (-2~~1+K~~0) direction, indicated by the thick blue line on the reciprocal space map in Fig.~3(d). Three energy transfers $\hbar\omega=2,6,$ and 10~meV are plotted, and there is no substantial change in the peak-width as a function of energy transfer. In Fig.~3(b), $S_{mag}(\vec{Q},\,\omega)$ is integrated over $1 \leq\hbar\omega\leq 11$~meV and compared to the calculated equal-time $S_{mag}(\vec{Q},\,\omega)$ for uncorrelated nearest-neighbor singlets.  The solid line corresponds to the result of the uncorrelated nearest-neighbor singlet model multiplied by $|F(\vec{Q})|^{2}$ where $F(\vec{Q})$ is the free Cu$^{2+}$ magnetic form factor.  Here, the measured data clearly indicate longer range correlations than the nearest neighbor singlet model.  Figure~3(c) depicts a line-scan of the dynamic structure factor (integrated over 1 $\leq\hbar\omega\leq$ 7~meV) along the $(0~K~0)$ direction. The nearest neighbor singlet model does not account for the observed scattering intensity at the $(0,2,0)$-type positions.

\

Further evidence of the continuum nature of the scattering is shown in Figure~4(a) where $S_{tot}(\vec{Q},\,\omega)$ is plotted along the $K-\Gamma-K$ direction in the $(1,0,0)$ Brillouin zone.  For 2 $\leq\hbar\omega\leq$ 7~meV, the scattered intensity is nearly constant along this direction. Also, the data shows another point of contrast to the nearest neighbor singlet calculation which predicts slightly larger intensities near the $K$ points. A recent theoretical calculation for $S_{mag}(\vec{Q},\,\omega)$ based on spinon excitations\cite{Hao} has a similar pattern as the singlet model with higher intensities near the $K$ points. The data show no evidence for significantly higher intensity near the $K$ points.  Indeed, as mentioned previously, at low energy transfers $\hbar\omega < 2$~meV the intensity is largest at the $\Gamma$ point, as shown in Fig.4(b) for $\hbar\omega = 0.75$~meV.

\

A central question for the ground state of herbertsmithite is whether a spin-gap exists. The answer is crucial for the classification of its ground-state.  One surprising aspect of our data is that the spin excitations appear to be gapless over a wide range of $\vec{Q}$ positions, at least down to $\hbar \omega = 0.25$~meV. This observation is difficult to reconcile with the ground-state properties of valence bond crystals\cite{SinghHuse} or gapped spin liquids.  Even when compared to theories for gapless spin liquids, most predict only a small set of special reciprocal lattice points for which the excitations are truly gapless.\cite{Ran,Fisher}  One possible caveat to our finding is that the small percentage of weakly interacting impurities in the interlayer sites may hide an intrinsic spin gap due to the kagom\'{e} spins.  However, it is likely that the impurities only affect the excitations below 1 meV where the upturn in intensity is seen with decreasing energy transfer.  Thus, the hexagonal ring pattern of the structure factor for $1.5 \leq\hbar\omega\leq 11$~meV is undoubtedly intrinsic to the kagom\'{e} layers. And consequently, this sets a conservative upper bound for the intrinsic spin-gap to be $\sim J/10$, if it exists.  Again, this applies for every $\vec{Q}$ position where the magnetic signal is seen. It may also be necessary for the theoretical calculations based on the Heisenberg model on the kagom\'{e} lattice to be modified to more closely match the spin Hamiltonian of herbertsmithite.

\

The observed spinon continuum is the strongest evidence yet that the ground state of the $S=\frac{1}{2}$ kagom\'{e} antiferromagnet herbertsmithite is a quantum spin liquid.  The data are consistent with a short-range RVB state, with spin correlations that go beyond nearest neighbors. An intriguing aspect of quantum spin liquids is that while the spin correlations may be short-ranged, the quantum coherence is long-ranged. These neutron results serve as a strong foundation for detailed tests of theoretical proposals for spin liquid states on the kagom\'{e} lattice.

\

\clearpage

\textbf{Figure 1.}  Intensity for inelastic neutron scattering from a single crystal sample of ZnCu$_3$(OD)$_6$Cl$_2$ measured at $T=6$~K.  The dynamic structure factor $S_{tot}(\vec{Q},\,\omega)$ is plotted for (a) $\hbar\omega=6$~meV and (b) $\hbar\omega=2$~meV with $E_{f}=5.1$~meV and (c) $\hbar\omega=0.75$~meV with $E_{f}=3.0$~meV.  A sample-out background was measured and subtracted.  The diffuse scattering is mostly magnetic in origin, since the phonon contribution to the signal is small (except near the $(2,2,0)$-type positions where the fundamental Bragg peaks are strong). (d) The magnetic part of the dynamic structure factor $S_{mag}(\vec{Q},\,\omega)$ integrated over 1 $\leq\hbar\omega\leq 9$~meV. (e) Calculation of the equal-time structure factor $S_{mag}(\vec{Q})$ for a model of uncorrelated nearest-neighbor dimers. The intensity corresponds to 1/8 of the total moment sum rule $S(S+1)$ for the spins on the kagom\'{e} lattice.  The data presented in parts (a)-(c) are normalized to Barns/(sr.~eV~form.unit) as shown by the left color bars. The data presented in parts (d)-(e) are unitless, with the scale given by the right color bar. The Brillouin zone boundaries are drawn in the figure for clarity; they correspond to the conventional unit cell with $a=b=6.83$~{\AA}, $c=14.05$~{\AA}, $\alpha=\beta=90^{\circ}$, and $\gamma=120^{\circ}$.

\

\

\textbf{Figure 2.} Intensity contour plots of the dynamic structure factor as a function of $\hbar\omega$ and $\vec{Q}$ for two high-symmetry directions (a) the $(H~0~0)$ direction and (b) the $(H~H~0)$ direction.  These directions are indicated by the thick black lines on the reciprocal space map shown in (d). Along the $(H~H~0)$ direction, a broad excitation continuum is observed over the entire range measured. The color bar shows the magnitude of $S_{tot}(\vec{Q},\,\omega)$ in Barns/(sr.~eV~form.unit). (c) Energy dependence of $S_{tot}(\vec{Q},\,\omega)$ measured at high symmetry reciprocal space locations. Data with $\hbar\omega\geq 1.5$~meV are taken with $E_{f}=5.1$~meV, whereas data with $\hbar\omega\leq 1$ meV are taken with $E_{f}=3.0$~meV for better energy resolution (except for at $\Gamma^{*}$, which are taken with $E_{f}=5.1$~meV). (Inset) Energy dependence of $S_{mag}(\vec{Q},\,\omega)$ where the non-magnetic scattering from the sample is subtracted. (d) Legend showing the integrated areas in reciprocal space referred to in the previous parts of the figure.

\

\

\textbf{Figure 3.} Plots of the dynamic structure factor along specific directions in reciprocal space. (a) $S_{mag}(\vec{Q},\,\omega)$ along the (-2~~1+K~~0) direction, indicated by the thick blue line on the reciprocal space map in part (d). Three energy transfers $\hbar\omega=2$,6, and 10~meV are shown.  (b) $S_{mag}(\vec{Q},\,\omega)$ along the (-2~~1+K~~0) direction integrated over 1 $\leq\hbar\omega\leq$ 11~meV. (c) $S_{mag}(\vec{Q},\,\omega)$ along the $(0~K~0)$ direction, indicated by the thick orange line on the reciprocal space map in part (d), integrated over 1 $\leq\hbar\omega\leq$ 7~meV. The solid lines in parts (b) and (c) are the calculated equal-time structure factor for uncorrelated nearest-neighbor dimers multiplied by $|F(\vec{Q})|^{2}$ where $F(\vec{Q})$ is the free Cu$^{2+}$ magnetic form factor. (d) Legend showing the trajectories in reciprocal space referred to in the previous parts.

\

\

\textbf{Figure 4.}  (a) Intensity contour plot of the dynamic structure factor as a function of $\hbar\omega$ and $\vec{Q}$ for the $(K-\Gamma-K)$ direction shown as the thick orange line in the inset. An excitation continuum is clearly observed.  The color bar indicates the magnitude of $S_{tot}(\vec{Q},\,\omega)$ in Barns/(sr.~eV~form.unit). (b) A line cut plotting $S_{mag}(\vec{Q},\,\omega)$ along the $(K-\Gamma-K)$ direction with $\hbar\omega=0.75$~meV measured with $E_{f}=3.0$~meV.

\clearpage

\begin{figure}[h!]
\includegraphics[width=12cm]{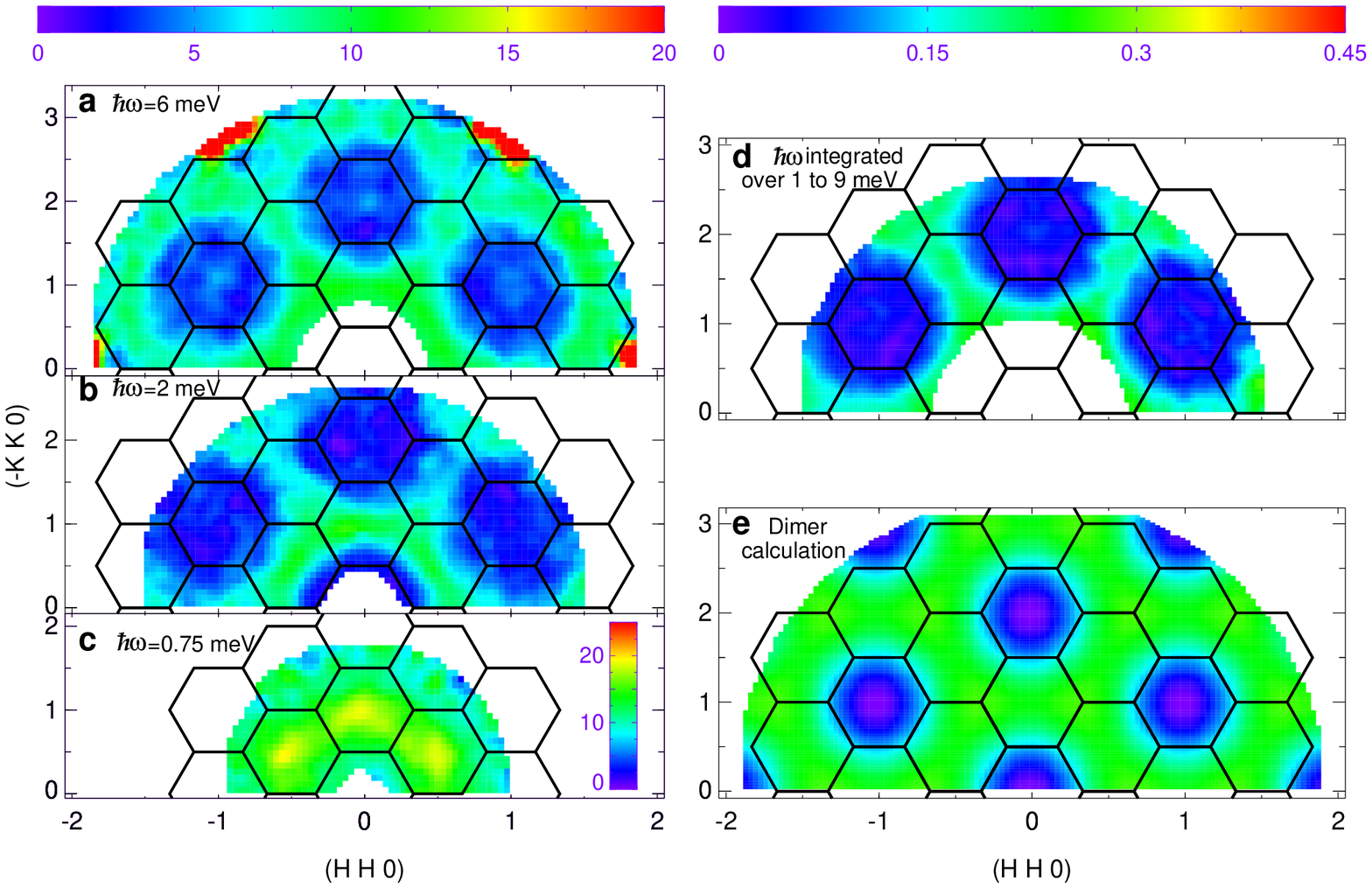} \vspace{0mm}
\caption{\label{Figure1}}
\end{figure}

\

\

\begin{figure}[h!]
\includegraphics[width=12cm]{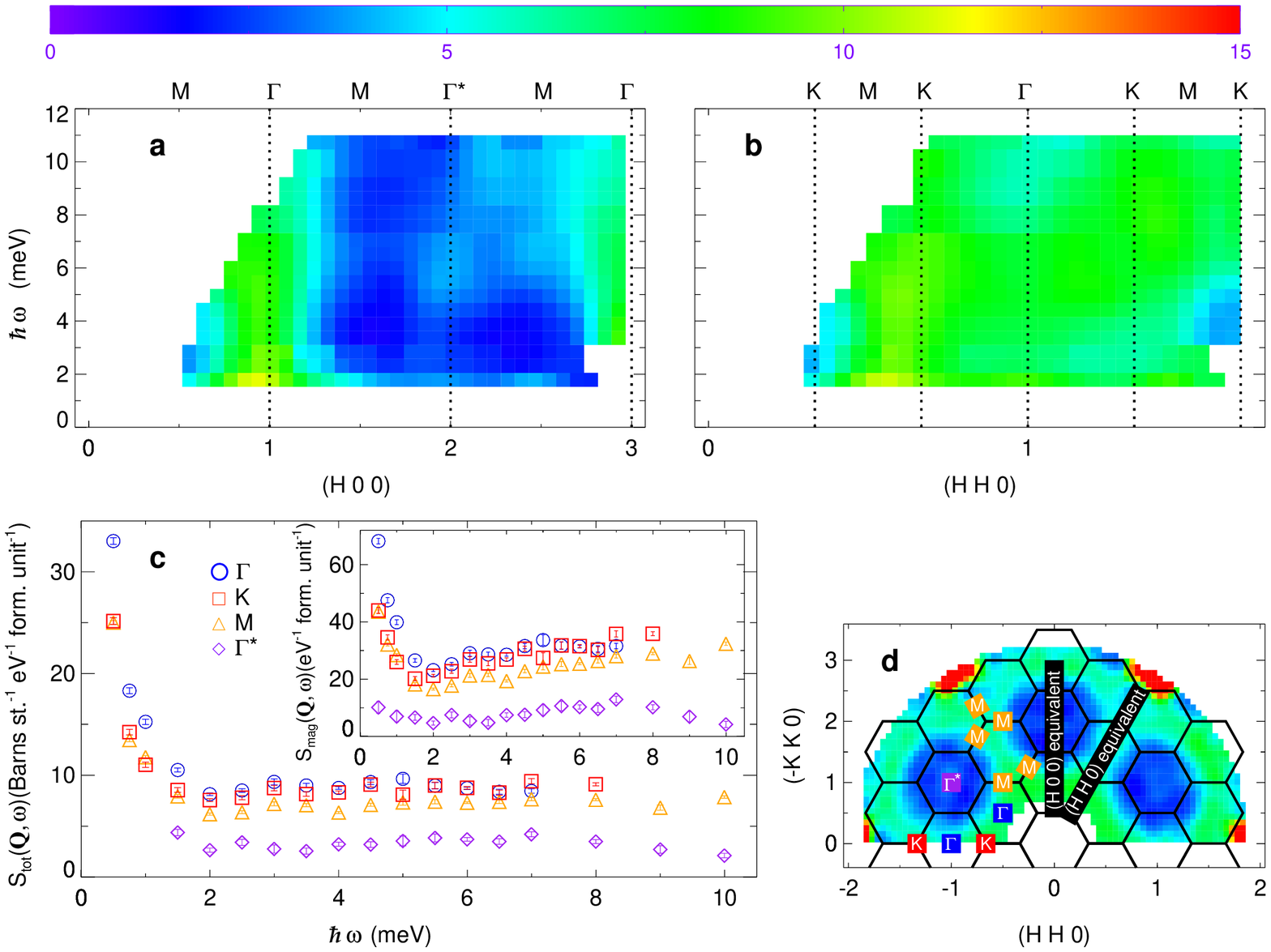} \vspace{0mm}
\caption{\label{Figure2}}
\end{figure}

\

\

\begin{figure}[h!]
\includegraphics[width=12cm]{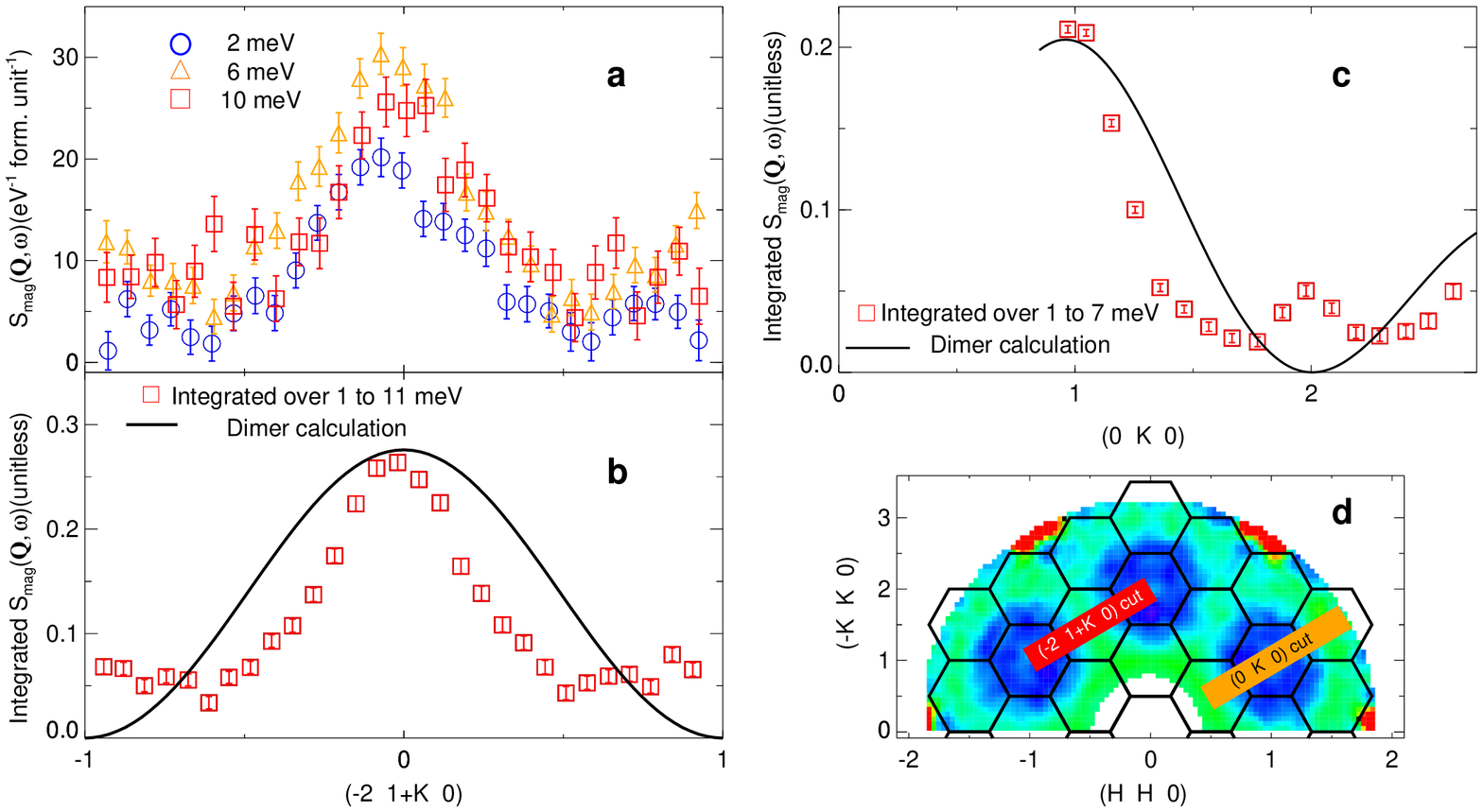} \vspace{0mm}
\caption{\label{Figure3}}
\end{figure}

\

\

\begin{figure}[h!]
\includegraphics[width=12cm]{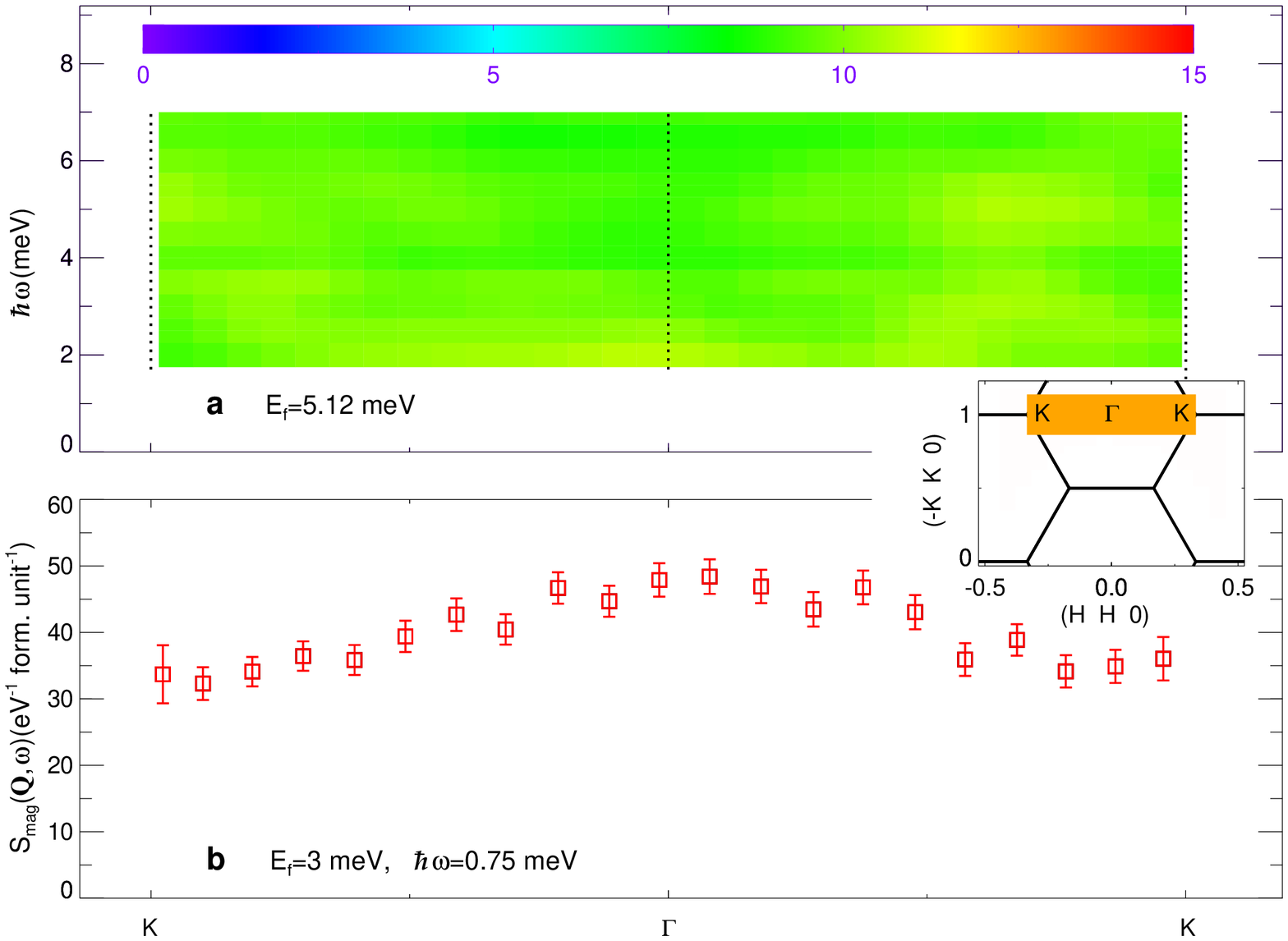} \vspace{0mm}
\caption{\label{Figure4}}
\end{figure}

\clearpage
\bibliography{Nature_v492_p406_2012}

\

\noindent{\textbf{Acknowledgements}}

We acknowledge T. Senthil, P.A. Lee, Z. Hao, and O. Tchernyshyov for valuable discussions, and J. Wen for assistance in data reduction. The work at MIT was supported by the Department of Energy (DOE), Office of Science, Office of Basic Energy Sciences (BES) under Grant No. DE-FG02-07ER46134.  This work utilized facilities supported in part by the National Science Foundation under Agreement No. DMR-0944772.  The work at IQM was supported by the DOE, Office of Basic Energy Sciences, Division of Material Sciences and Engineering under Award DE-FG02-08ER46544.

\

\noindent{\textbf{Author contributions}}

Y.S.L. supervised all aspects of the research.  T.-H.H. synthesized and characterized the sample. T.-H.H., J.A.R., J.S.H., and C.B. collected the neutron scattering data. T.-H.H. analyzed the data with input from Y.S.L., C.B., J.S.H, and J.A.R. S.C. and D.G.N. aided in the sample preparation process.  T.-H.H and Y.S.L. wrote the manuscript with comments from all others.  The manuscript reflects the contributions of all authors.

\

Correspondence and requests for materials should be addressed to younglee@mit.edu or tianheng@alum.mit.edu.

\

$^{\dag}$Present address: The James Franck Institute and Department of Physics, The University of Chicago, Chicago, Illinois 60637, USA and
Materials Science Division, Argonne National Laboratory, Argonne, Illnois 60439, USA. \\

\clearpage

\setcounter{figure}{0}

\noindent {\large{\textbf{Supplementary Information}}}

\subsection{Single crystal growth}

Large single crystals (up to $\sim 1$ cm in linear dimension) of deuterated herbertsmithite (ZnCu$_{3}$(OD)$_{6}$Cl$_{2}$) were grown with a technique similar to that published for non-deuterated crystals \cite{Han}.  A picture of one resulting crystal weighing 0.32 gram is shown in Fig.~S1(a).  A quartz tube (12.7~mm~inner diameter, 19.1~mm~outer diameter) was charged with starting materials CuO (1.65~g, 20.7~mmol), ZnCl$_{2}$ (17.0~g, 125~mmol), and D$_{2}$O (31.0~g, 28.0~mL).  The tube was purged of air with a mechanical pump and then sealed.  The mixture was pre-reacted in a box furnace for two days at 185$^{\circ}$C, producing a green-blue ZnCu$_{3}$(OD)$_{6}$Cl$_{2}$ microcrystalline powder. The powder was collected at one end of the tube, which was then placed horizontally inside of a three-zone furnace.  The temperature of all three zones was slowly raised to 180$^{\circ}$C; a temperature gradient was then applied to allow for crystallization transport across the tube. After approximately ten months, all of the powder had been transported through the solution and crystallized at the cold end of the tube, yielding multiple large single crystals.  The temperature gradient near the cold end was measured to be 1$^{\circ}$C/cm.

\

The crystals were characterized using chemical analysis, single-crystal x-ray diffraction, and magnetization measurements, and their properties were found to be consistent with previous measurements of herbertsmithite.\cite{Han}  A schematic of the crystal structure is shown in Fig.~S1(b).  Fifteen of the largest single crystals were co-aligned on an aluminum sample holder, yielding a total mass of 1.2 grams for the herbertsmithite sample.  The overall sample mosaic was determined by neutron diffraction to be $\sim 2^{\circ}$. An identical aluminum sample holder was also prepared for the purpose of background subtraction. No decomposition of the crystals has been observed in air, water, acetone, or paratone oil. The sample was stored in a can sealed with gaseous helium when not used for measurements.

\subsection{Data analysis}

The scattering cross section measured during an experiment has three contributions: the magnetic scattering from the sample, the nuclear scattering from the sample (including the nuclear incoherent scattering and the coherent scattering from Bragg peaks and phonons), and the background count rate (including scattering from the sample holder and the sample environment). The latter background was measured with the empty sample holder inside the cryostat for each instrumental configuration used (that is, for all measured energy transfers $\hbar\omega$ with either $E_{f}=3$~meV and 5.12~meV), and subtracted from the corresponding sample-in data.  For all data sets, a monitor correction factor was applied which takes into account the $\lambda/2$ neutron contributions to the monitor counts.  The error bars correspond to standard statistical uncertainties.

\

The nuclear (non-magnetic) scattering from the sample can be estimated by comparing the data at low energy transfers with similar data measured previously on a powder sample \cite{Helton2}. For an energy transfer of $\hbar\omega=0.5$~meV (with $E_{f}=3$~meV), the measured intensities at $T=1.6$~K and 60~K were integrated over a substantial region of reciprocal space ($0.6 < |\vec{Q}| < 1.6 {\AA}^{-1}$) to deduce the local susceptibility. The data taken from the empty sample holder at $T=1.6$~K, integrated over the same $|\vec{Q}|$ range, was subtracted at both temperatures. The $|\vec{Q}|$-integrated magnetic cross section from the single crystal sample is assumed to follow the same temperature dependence as measured in powder samples \cite{Helton2} at $T=1.6$~K and 60~K. Also assuming that the non-magnetic scattering (predominately incoherent elastic scattering) from the sample is temperature-independent in this temperature range, the magnetic and the non-magnetic contributions to the scattering at $\hbar\omega=0.5$~meV and $T=1.6$~K can be obtained. In a similar manner, the magnetic and the non-magnetic contributions can be calculated for each $\hbar\omega$ with $E_{f}=3$~meV at $T=1.6$~K. Finally, the separation of the magnetic and non-magnetic contributions for $E_{f}=5.12$~meV and $\hbar\omega\leq$2~meV can be achieved using these results along with knowledge of the energy resolution function. The contribution of non-magnetic scattering from the sample for $\hbar\omega\geq 2$~meV is fixed at the value it has at $\hbar\omega=2$~meV.  After subtracting the estimated non-magnetic scattering from the sample, one obtains the magnetic part of the dynamic structure factor $S_{mag}(\vec{Q},\,\omega)$, where $S_{mag}(\vec{Q},\,\omega)=\frac{1}{2\pi}\int^\infty_{-\infty} dt \,\, e^{-i\omega t} \sum_{\vec{r}} \, e^{i \vec{Q} \cdot \vec{r}} \,\, \langle S^\alpha_0(0) \, S^\beta_{\vec{r}}(t) \rangle$, where the angle brackets denote an average over configurations and $\alpha,\beta$ refer to $x,y,z$ vector components.

\subsection{Additional data}

The measured scattering broadens slightly upon warming to $T=125$~K, but the basic diffuse pattern remains. Figure~S2(a) shows the data for $\hbar\omega=6$~meV and $T=125$~K.  Here, the phonon background due to the empty sample holder was measured at $T=1.6$~K with the same instrumental configuration and then scaled to $T=125$~K using the thermal factor $n(\omega,T)+1$ ($n(\omega,T)$ is the Bose occupancy factor) and subtracted. For comparison, the low temperature $T=1.6$~K scattering for $\hbar\omega=10$~meV is shown in Fig.~S2(b), where the diffuse pattern is similar to the data for lower $\hbar\omega$ previously shown in Figure 1 of the letter.  Another way to directly compare the temperature dependence is to subtract the differential cross section measured at $T=125$~K from that measured at $T=1.6$~K (which yields the quantity $S_{tot 1.6K}(\vec{Q},\,\omega)- S_{tot               125K}(\vec{Q},\,\omega)$). This quantity is plotted in Figures~S2(c)-(e) for $\hbar\omega=2$~meV, 6meV, and 10meV.  At small wavevectors (where the phonon scattering is weak), the intensity is nearly temperature-independent up to $T=125$~K. However, we note that the intensity near (2 0 0) and equivalent positions appears to be slightly enhanced at higher temperatures.

\

Fits to the line-scans of $S_{mag}(\vec{Q},\,\omega)$ along the (-2~~1+K~~0) direction are shown in Figure~S3(a)-(c).
The fitted line shapes denote Lorentzians, and the estimated magnetic correlation lengths of about 3.1~{\AA} for each energy transfer are determined from 1/HWHM. A fit to the line-scan of $S_{mag}(\vec{Q},\,\omega)$ along the longitudinal (0~K~0) direction is shown in Fig.~S3(d). The fitted line shape denotes a Lorentzian, yielding a slightly larger estimated magnetic correlation length of about 5.1~{\AA}.  We note that the sample mosaic does not contribute to the reciprocal space width in this latter direction.  These estimates of the correlation length are comparable to the distance between Cu$^{2+}$ ions.

\vspace{5cm}

\renewcommand{\thefigure}{S\arabic{figure}}

\begin{figure}[h!]
\centering
\includegraphics[width=9cm]{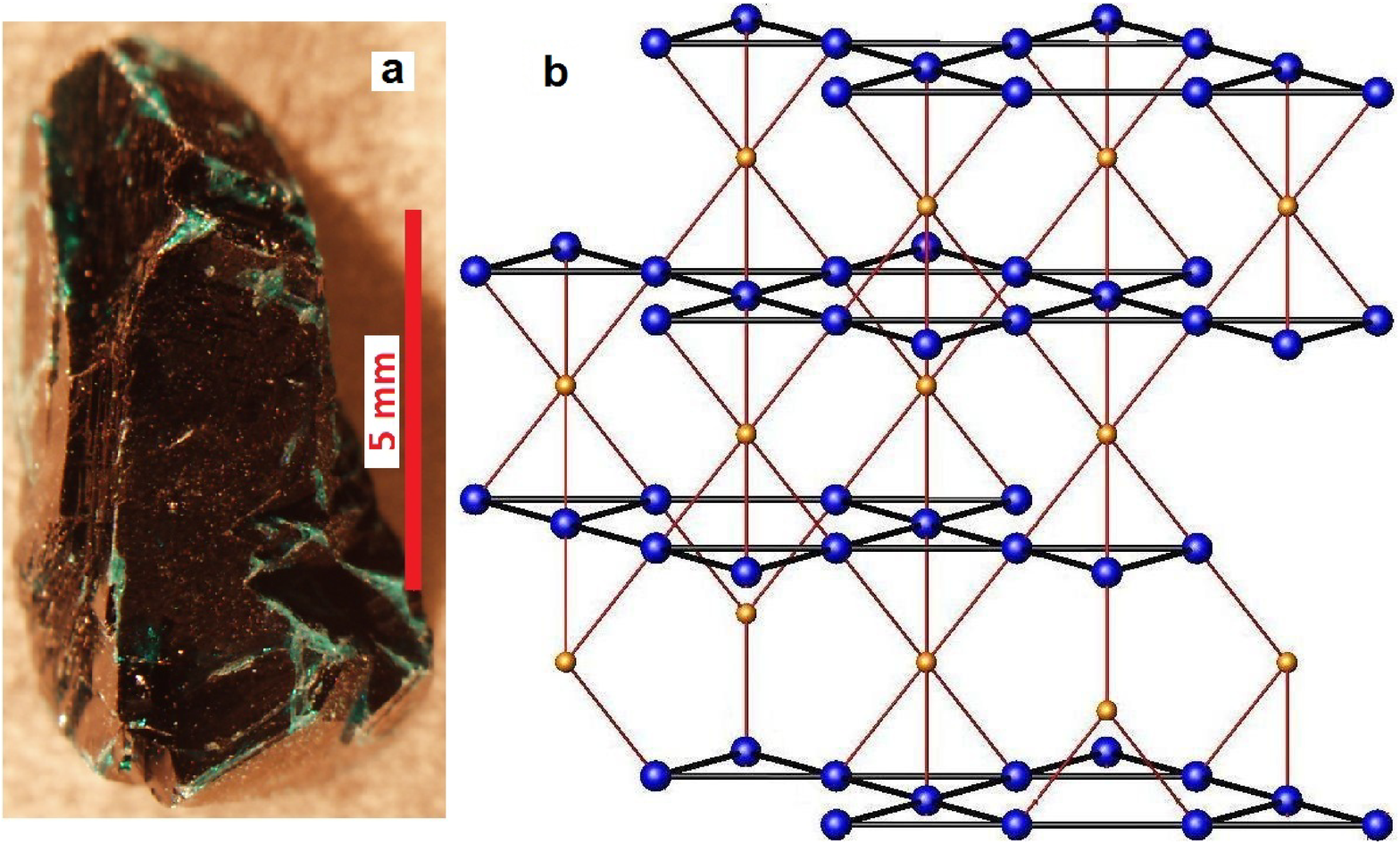} \vspace{-4mm}
\caption{(a) A large single crystal of herbertsmithite (one of the fifteen pieces co-aligned for the inelastic neutron scattering measurements). (b) Structure of ZnCu$_{3}$(OD)$_{6}$Cl$_{2}$ with only Cu$^{2+}$ ions (large blue spheres) and Zn$^{2+}$ ions (small brown spheres) displayed. The Cu-Cu bonds (thick black lines) are all equivalent
as are the Cu-Zn bonds (thin red lines).}
\end{figure}

\

\begin{figure}[h!]
\centering
\includegraphics[width=12cm]{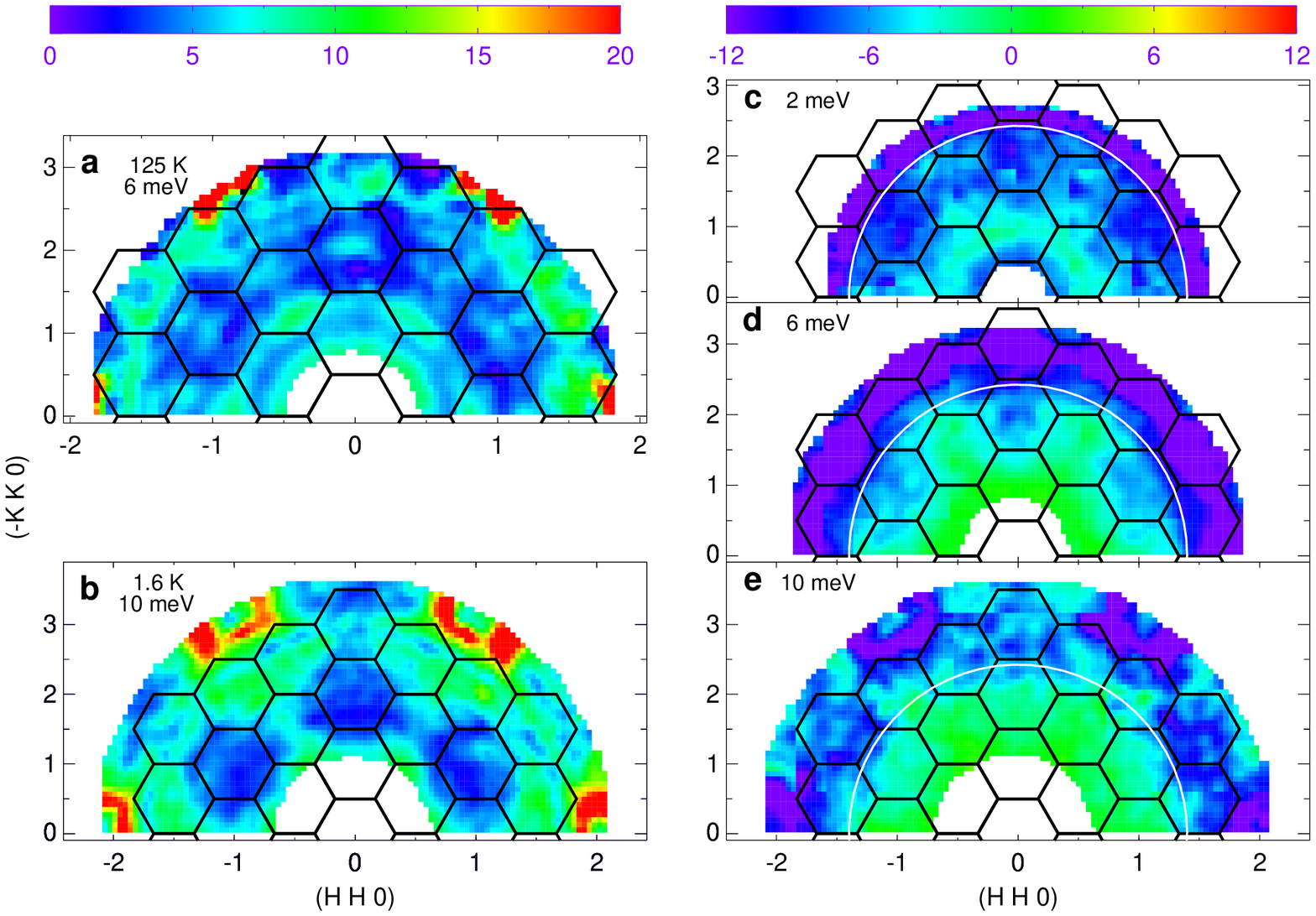} \vspace{-4mm}
\caption{(a) Contour plot of the dynamic structure factor $S_{tot}(\vec{Q},\,\omega)$ measured at $T=125$~K and $\hbar\omega=6$~meV (with $E_{f}=5.12$~meV). (b) Contour plot of $S_{tot}(\vec{Q},\,\omega)$ measured at $T=1.6$~K and $\hbar\omega=10$~meV (with $E_{f}$=5.12~meV). The difference between the dynamic structure factor measured at $T=125$~K from that measured at $T=1.6$~K ($S_{tot 1.6K}(\vec{Q},\,\omega) - S_{tot               125K}(\vec{Q},\,\omega)$) for (c) $\hbar\omega=2$~meV, (d) $\hbar\omega=6$~meV, and (e) $\hbar\omega=10$~meV. The areas outside the white circular lines are strongly affected by phonon scattering from the aluminum in the sample holder. The left color bar is for parts (a)-(b) and the right color bar is for parts (c)-(e). Both color bars are normalized in Barns sr.$^{-1} eV^{-1}$ form.unit$^{-1}$.}                                                                                                                                                      \end{figure}

\

\begin{figure}[h!]                                                                                                                                                        \centering                                                                                                                                                             \includegraphics[width=12cm]{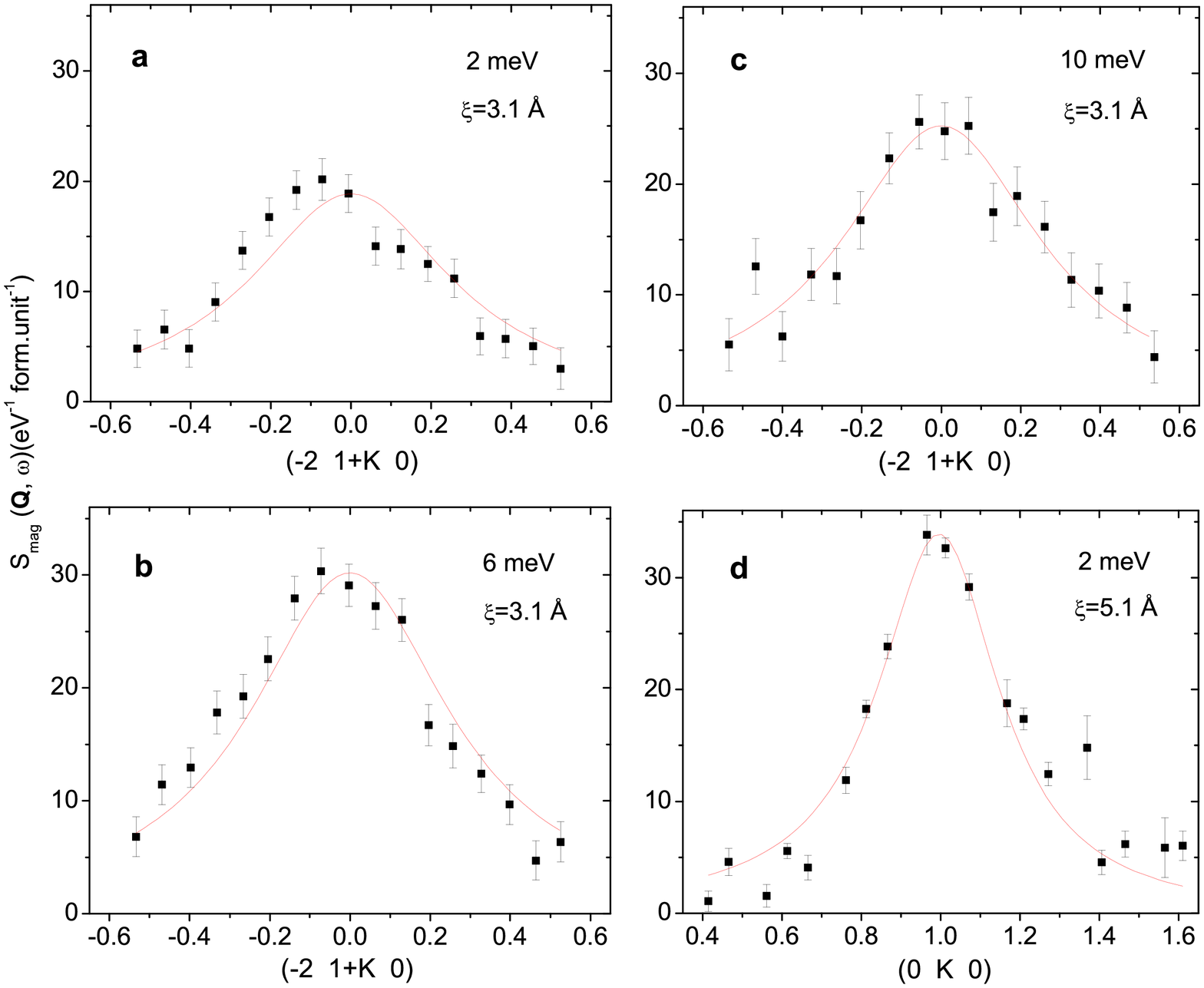} \vspace{-4mm}                                                                                                                \caption{ Line-scans of $S_{mag}(\vec{Q},\,\omega)$ along the (-2~~1+K~~0) direction for (a) $\hbar\omega=2$~meV, (b) $\hbar\omega=6$~meV, and (c) $\hbar\omega=10$~meV.  The solid lines denote Lorentzian fits, and the spin correlation length shown is estimated using 1/HWHM of the fitted peak. (d) Line-scan of $S_{mag}(\vec{Q},\,\omega)$ along the (0~K~0) direction at $\hbar\omega=2$~meV where the solid line denotes a Lorentzian fit.}
\end{figure}

\end{document}